\documentclass[sigconf,natbib=true]{acmart}
\usepackage[normalem]{ulem}
\usepackage{algorithm}
\usepackage{algpseudocode}
\usepackage{amsmath}
\usepackage{subfig}
\usepackage{diagbox}

\AtBeginDocument{%
  }

\copyrightyear{2023}
\acmYear{2023}
\setcopyright{acmlicensed}\acmConference[SIGIR '23]{Proceedings of the 46th International ACM SIGIR Conference on Research and Development in Information Retrieval}{July 23--27, 2023}{Taipei, Taiwan}
\acmBooktitle{Proceedings of the 46th International ACM SIGIR Conference on Research and Development in Information Retrieval (SIGIR '23), July 23--27, 2023, Taipei, Taiwan}
\acmPrice{15.00}
\acmDOI{10.1145/3539618.3591722}
\acmISBN{978-1-4503-9408-6/23/07}
\begin{document}

\title{Manipulating Federated Recommender Systems: Poisoning with Synthetic Users and Its Countermeasures}

\author{Wei Yuan}
\affiliation{%
  \institution{The University of Queensland}
  \city{Brisbane}
  \country{Australia}
}
\email{w.yuan@uq.edu.au}
\author{Quoc Viet Hung Nguyen}
\affiliation{%
  \institution{Griffith University}
  \city{Gold Coast}
  \country{Australia}
}
\email{henry.nguyen@griffith.edu.au}

\author{Tieke He}
\affiliation{%
  \institution{Nanjing University}
  \city{Nanjing}
  \country{China}}
\email{hetieke@nju.edu.cn}

\author{Liang Chen}
\affiliation{%
  \institution{Sun Yat-Sen University}
  \city{Guangzhou}
  \country{China}
}
\email{chenliang6@mail.sysu.edu.cn}

\author{Hongzhi Yin}\authornote{Corresponding author.}
\affiliation{%
 \institution{The University of Queensland}
 \city{Brisbane}
 \country{Australia}}
\email{db.hongzhi@gmail.com}
\renewcommand{\shortauthors}{Wei Yuan, Quoc Viet Hung Nguyen, Tieke He, Liang Chen, \& Hongzhi Yin}

\begin{abstract}
  Federated Recommender Systems (FedRecs) are considered privacy-preserving techniques to collaboratively learn a recommendation model without sharing user data.
  Since all participants can directly influence the systems by uploading gradients, FedRecs are vulnerable to poisoning attacks of malicious clients.
  However, most existing poisoning attacks on FedRecs are either based on some prior knowledge or with less effectiveness.
  To reveal the real vulnerability of FedRecs, in this paper, we present a new poisoning attack method to manipulate target items' ranks and exposure rates effectively in the top-$K$ recommendation without 
  relying on any prior knowledge.
  Specifically, our attack manipulates target items' exposure rate by a group of synthetic malicious users who upload poisoned gradients considering target items' alternative products.
  We conduct extensive experiments with two widely used FedRecs (Fed-NCF and Fed-LightGCN) on two real-world recommendation datasets.
  The experimental results show that our attack can significantly improve the exposure rate of unpopular target items with extremely fewer malicious users and fewer global epochs than state-of-the-art attacks.
  In addition to disclosing the security hole, we design a novel countermeasure for poisoning attacks on FedRecs. 
  Specifically, we propose a hierarchical gradient clipping with sparsified updating to defend against existing poisoning attacks.
  The empirical results demonstrate that the proposed defending mechanism improves the robustness of FedRecs.

\end{abstract}




\maketitle

\section{Introduction}~\label{sec_intro}
Recommender systems are widely deployed in many online scenarios (e.g., e-commerce~\cite{wei2007survey,chen2020try} and social media~\cite{yin2015joint,yin2016spatio,wang2020next}) to help users discover what they are interested in  from mass information.
Traditional recommender systems require collecting users' personal data to train recommendation models in a centralized way~\cite{zhang2019deep}. 
With the growing concerns about privacy and the new privacy protection regulations (e.g. GDPR~\cite{voigt2017eu} in the EU and CCPA~\cite{harding2019understanding} in the USA), 
federated recommenders (FedRecs)~\cite{ammad2019federated} have recently emerged as a privacy-preserving solution to collaboratively learn a recommendation model among personal devices without uploading users' raw data to a central server.

Although FedRecs can alleviate the privacy concerns of training recommender systems, recent studies~\cite{zhang2022pipattack,rong2022fedrecattack,rong2022poisoning} show that FedRecs are inherently vulnerable to gradient poisoning attacks (also called model poisoning attacks) as their open and decentralized characteristics allow any client to participate in the training process, and malicious clients may get involved~\cite{tolpegin2020data}.
Specifically, users with compromised devices will upload polluted gradients to achieve targeted adversarial goals (e.g. manipulate item rank and exposure rate).
PipAttack~\cite{zhang2022pipattack} presents the first work of gradient poisoning attacks on FedRecs to promote target items' exposure chances.
But it requires many malicious users and assumes adversaries know all items' popularity information.
FedRecAttack~\cite{rong2022fedrecattack} achieves item promotion with fewer malicious users. However, it is based on a stronger assumption that malicious users can acquire a proportion of data from benign users, which is not applicable in most FedRecs.
~\cite{rong2022poisoning} proposed a gradient poisoning attack on FedRecs without prior knowledge assumption. Nevertheless, its performance is unstable and undesirable because it randomly samples vectors from a Gaussian distribution to act as the proxy of the target item's embedding.
To sum up, existing gradient poisoning attacks on FedRecs are either based on impractical prior knowledge, or with undesirable performance.
Furthermore, although these works highlight the urgent need for new defense mechanisms against gradient attacks, no effective defense solution has been developed.

To disclose the real threats of poisoning attacks to FedRecs,  we present a new gradient poisoning attack method named ``PSMU'' (\emph{P}oisoning with \emph{S}ynthetic \emph{M}alicious \emph{U}sers) in this paper.
PSMU aims to improve target items' exposure rate (i.e., to make target items  appear in more users' top-K recommendation lists) by generating and uploading polluted gradients with a group of malicious users.
The idea of PSMU is based on our interesting finding that the similarity between the top-K recommendations of randomly constructed synthetic users and real users is surprisingly high due to popularity bias~\cite{abdollahpouri2019managing}, as shown in Table 1. Based on this finding, we propose an assumption that if a target item has a high exposure rate among synthetic users, then this item will have a high chance of being recommended to real users by FedRecs. With this assumption, PSMU works as follows. A malicious user randomly selects a set of items as interacted items when participating in FedRec's training process.
Then, the malicious user learns a synthetic user embedding based on the randomly selected positive items and optimizes the target item's rank based on the synthetic user embedding.
Besides, to further improve the target items' competition,  we enlarge the competition set by adding the alternatives of the target items. The predicted preference scores of the target items are required to be higher than that of top-K recommended items and their substitute items (i.e., the competition set of target items).

The vulnerability of FedRecs highlights the timely demand for a new defense against poisoning attacks.
Alas, no existing work attempts to provide a solution for this security issue in FedRecs.
Some research has been done in general federated learning (e.g. federated classification~\cite{zhang2021survey}), but it cannot be directly applied to FedRecs because of the following major differences between FedRecs and general federated learning.
(1) Unlike general federated learning, the data from different clients are not IID. Therefore, the same item's gradients from different clients may vary significantly from each other.
However, the widely used Byzantine defense methods (e.g. Krum~\cite{blanchard2017machine}, Bulyan~\cite{guerraoui2018hidden}, Trimmed Mean~\cite{yin2018byzantine}) in federated learning generally assume the clients' data are with the same distribution, and directly compare clients' uploaded gradients to eliminate poisoning effects. Therefore, they usually cause a significant performance drop in FedRecs (see details in Section~\ref{sec_rq3}).
(2) Compared with general federated learning, the server in FedRecs cannot access clients' private parameters. 
Therefore, existing defense methods relying on accessing the whole model  cannot work in FedRecs~\cite{fang2020local}.

In this paper, we propose a novel defense method against gradient poisoning attacks at the central server, named \emph{hi}erarchical gradient \emph{c}lipping with \emph{s}parsified updating (HiCS).
At first, the central server clips all received gradients to avoid dominated gradients.
The first-clip limits poisoned gradients' effects. However, it is still insufficient since we cannot set a too-small clipping factor to guarantee convergence.
Thus, an adaptive clipping with sparsified updating is further employed against gradient poisoning attacks.
Specifically, the clipped gradients will accumulate in a memory bank.
The server only selects several most significant item embedding gradients in the bank to update the model.
Before updating, another clipping is applied to the accumulated gradients with an adaptive clipping factor to further reduce the polluted gradients' influence.

To demonstrate the generalization and effectiveness of our proposed attack and defense, we conduct extensive experiments with two commonly used FedRecs (Fed-NCF~\cite{ammad2019federated} and Fed-LightGCN~\cite{he2020lightgcn}) on two real-world datasets (MovieLens-1M~\cite{harper2015movielens} and Amazon Digital Music~\cite{mcauley2015image}).
The experimental results validate the threats of gradient poisoning attacks on FedRecs, even without any prior knowledge of users and items and with extremely fewer malicious users. Meanwhile, the results also show the effectiveness of our proposed defense method against all existing poisoned gradient attacks on FedRecs.

In conclusion, the main contributions of this paper are as follows:
\begin{itemize}
 \item We present an interesting finding that  there is a large portion of overlapped items   between the top-K recommendations of randomly constructed synthetic users and real users, which strongly supports our conjecture that if a target item enjoys a high exposure rate among the synthetic users, it will have a high exposure rate among real users in CF-based recommendation systems. 
  \item We present an effective gradient poisoning attack method for FedRecs, namely PSMU, which can manipulate items' ranks without prior knowledge and with much fewer malicious users and fewer global epochs, disclosing more severe vulnerability of FedRecs to gradient poisoning attacks.
  \item To the best of our knowledge, we are the first to propose a defense method (HiCS) based on gradient clipping and sparsified updating to address the threats of gradient poisoning attacks on FedRecs.
  \item Extensive experiments are conducted with two widely used FedRecs on two real-world recommendation datasets, validating the generalization and effectiveness of our attack and defense methods. 
\end{itemize}

\section{Related Work}
\subsection{Federated Recommender Systems}
Since FedRecs provide a privacy-preserving solution to train a recommender system~\cite{zheng2016keyword}, they attract increasing attention in recent years.
Ammad et al.~\cite{ammad2019federated} provided the first FedRec framework.
After that, many extended versions are proposed in recent years~\cite{long2023decentralized}.
Muhammad et al.~\cite{muhammad2020fedfast} ameliorated user sampling and aggregation strategy to accelerate FedRec's convergency process.
FedRec++~\cite{liang2021fedrec++} explores a secure way to learn recommender with explicit feedback.
~\cite{wu2021fedgnn} and~\cite{wu2022fedcl} attempt to use Graph Neural Networks~\cite{scarselli2008graph} and Contrastive Learning~\cite{chen2020simple,yu2022graph} in FedRecs.
Imran et al.~\cite{imran2022refrs} proposed a resource-efficient FedRec to learn user preferences.

\subsection{Attacks on Federated Recommender Systems}
With the wide application of FedRecs, the potential security issues raise researchers' concerns~\cite{zhang2022comprehensive,wang2022fast,yuan2023interaction},
and poisoning attack is one of the recently emerging threats.
In general, poisoning attacks can be classified into data poisoning and gradient poisoning.

\textbf{Data Poisoning Attack.} Data poisoning attack conducts attacks by injecting adversarial interactions to pollute the training data~\cite{kapoor2017review}.
To create high-quality fake interactions, data poisoning attacks have to utilize the whole dataset's information~\cite{fang2018poisoning,zhang2021data,huang2021data,fan2022survey,zhang2022targeted}.
Therefore, data poisoning attacks are usually studied in centralized recommender systems.
~\cite{wu2022fedattack} is the only data poisoning attack in FedRecs, but it focuses on reducing the accuracy of FedRecs, which is not related to our work's topic.

\textbf{Gradient Poisoning Attack.} Gradient poisoning attack (also called model poisoning attack) is specially targeted at Federated Learning (FL) scenarios~\cite{fang2020local}.
PipAttack~\cite{zhang2022pipattack} is the first work that conducts gradient poisoning attacks in FedRecs.
It promotes target items by adjusting their embeddings to be similar to popular items' embeddings.
PipAttack has two drawbacks: (1) it assumes that all items' popularity information is available for malicious users; (2) it relies on a large proportion of malicious users (e.g. more than $10\%$).
FedRecAttack~\cite{rong2022fedrecattack} achieves the attacker's goal with fewer malicious users, however, it is based on a stronger assumption that even breaks FedRec protocol, i.e., it requires accessing a proportion of interaction data from benign users.
~\cite{rong2022poisoning} is the only gradient poisoning attack that does not rely on any prior knowledge.
However, it is not effective enough since it simply approximates user embeddings based on randomly generated vectors from Gaussian distribution.
As a result, existing poisoning attacks are either requiring unobtainable prior knowledge or having ineffective performance, which cannot reveal the real threats of poisoning attacks to FedRecs.

\textbf{Defense.} The defense against poisoning attacks in FedRecs is still under-explored.
FRU~\cite{yuan2023federated} provides a recovery way by using federated unlearning~\cite{nguyen2022survey} to efficiently reconstruct destroyed FedRecs.
However, it cannot directly defend against poisoning attacks.

\section{Preliminaries}
In this section, we present preliminaries related to our research.
Note that the bold lowercase (e.g. $\mathbf{a}$) represents vectors, the bold uppercase (e.g. $\mathbf{A}$) means matrices, and the squiggle uppercase (e.g. $\mathcal{A}$) denotes sets.

\subsection{Federated Recommendation Framework}
Following~\cite{zhang2022pipattack,rong2022fedrecattack,rong2022poisoning}, we employ the most commonly used federated recommendation framework proposed by~\cite{ammad2019federated} as our base FedRec framework.
Most existing FedRecs~\cite{yang2020federated,imran2022refrs,muhammad2020fedfast,wang2022fast} are based on this framework.

Let $\mathcal{U}$ and $\mathcal{V}$ denote the sets of benign users (clients)\footnote{In this paper, client and user are equivalent, since a client is responsible for one user.} and items, respectively.
$\left| \mathcal{U}\right|$ and $\left|\mathcal{V}\right|$ are users' and items' sizes.
In FedRec, each user $u_{i}$ is a client who manages its' local training dataset $\mathcal{D}_{i}$.
$\mathcal{D}_{i}$ consists of many user-item interactions $(u_{i}, v_{j}, r_{ij})$, where $r_{ij}$ is a binary rating denoting whether user $u_{i}$ has interacted with item $v_{j}$. i.e., $r_{ij}=1$ means $u_{i}$ has interacted with $v_{j}$, while $r_{ij}=0$ indicates no interaction between $u_i$ and $v_j$.
$\mathcal{V}_{i}^{+}$ and $\mathcal{V}_{i}^{-}$ are used to denote the sets of interacted items and non-interacted items for user $u_{i}$.
The FedRec aims to predict $\hat{r}_{ij}$ between $u_{i}$ and each non-interacted item $v_j$ and recommend the top-K ones with the highest predicted scores.

In FedRec, a central server coordinates a large number of clients and the parameters of the recommender system can be divided into public and private parameters.
Private parameters are user embeddings $\mathbf{U}$, which are stored and maintained on users' devices locally.
For public parameters, we use $\mathbf{V}$ to denote item embeddings and $\mathbf{\Theta}$ represents all other public parameters such as layer weights.

\textbf{Federated Learning Protocol.} 
In the beginning, the central server initializes all public parameters and all clients initialize their corresponding private parameters.
Then, a recommender model is trained within the following steps at every global epoch.

At first, the central server randomly selects a subset of users $\mathcal{U}_{t-1}$ to participate in the training process and dispenses the public parameters $\mathbf{\Theta}^{t-1}$ and $\mathbf{V}^{t-1}$ to these users.
The selected users combine received public parameters and their local private parameters $\mathbf{u}_{i}^{t-1}$ to form a local recommender.
Then, the local recommender is updated on local dataset $\mathcal{D}_{i}$ by optimizing the loss function:
\begin{equation}\label{eq_ori_loss}
  \mathcal{L}^{rec} = -\sum\nolimits_{(u_{i}, v_{j}, r_{ij})\in \mathcal{D}_{i}} r_{ij}\log \hat{r}_{ij} + (1-r_{ij})\log (1-\hat{r}_{ij})
\end{equation}
After several local epochs of training, the selected user $u_{i}$ updates its private parameters as follows:
\begin{equation}\label{eq_local_update}
  \mathbf{u}_{i}^{t} = \mathbf{u}_{i}^{t-1} - lr\nabla \mathbf{u}_{i}^{t-1}
\end{equation}
where $lr$ is the learning rate.
The gradients of public parameters $\nabla \mathbf{\Theta}^{t-1}_{i}$ and $\nabla \mathbf{V}^{t-1}_{i}$ are uploaded to the central server.
The central server aggregates all the uploaded gradients to update the public parameters:
\begin{equation}\label{eq_aggregate}
  \begin{aligned}
    &\mathbf{V}^{t} = \mathbf{V}^{t-1} - lr\sum\limits_{u_{i}\in \mathcal{U}_{t-1}} \nabla \mathbf{V}^{t-1}_{i}\\
    &\mathbf{\Theta}^{t} = \mathbf{\Theta}^{t-1} - lr\sum\limits_{u_{i}\in \mathcal{U}_{t-1}} \nabla \mathbf{\Theta}^{t-1}_{i}  
  \end{aligned}
\end{equation}

\subsection{Base Federated Recommenders}
Neural Collaborative Filtering (NCF)~\cite{he2017neural} and LightGCN~\cite{he2020lightgcn} are two classical and widely used recommenders.
To show the generalization of our proposed attack and defense methods, 
we extend NCF and LightGCN with the above federated recommendation framework to form Fed-NCF and Fed-LightGCN.
Note that in Fed-LightGCN, the propagation is limited to the local bipartite graph on each client according to privacy-protection requirements.



\subsection{Gradient Poisoning Attack and Defense}
\textbf{Attack Goal.}
This paper focuses on targeted attacks on federated recommenders that aim to \emph{promote} target items $\widetilde{\mathcal{V}}$ to as many users as possible, which is the most common goal setting in poisoning attacks.
Specifically, given that a recommender system recommends $K$ items $\hat{\mathcal{V}}_{i}$ to user $u_{i}$, the goal of our attack is to improve each target item's Exposure Ratio at rank $K$ (ER@K)~\cite{zhang2022pipattack} defined as follows:
\begin{equation}
  \label{eq_er}
  ER@K = \frac{1}{\left|\widetilde{\mathcal{V}} \right|} \sum\limits_{v_{j} \in \widetilde{\mathcal{V}}} \frac{\left|\left\{ u_{i} \in \mathcal{U} | v_{j}\in \hat{\mathcal{V}}_{i} \land v_{j}\in \mathcal{V}_{i}^{-} \right\}\right|}{\left|\left\{ u_{i} \in \mathcal{U} | v_{j}\in \mathcal{V}_{i}^{-} \right\} \right|} 
\end{equation}
Besides, the attack does not significantly hurt the recommender's performance in order to keep the stealthiness and effectiveness of item promotion.

\textbf{Attack Approach.}
Given a set of malicious users $\widetilde{\mathcal{U}}$ in our attack, they need to poison gradients $\nabla \widetilde{\mathbf{\Theta}}$ and $\nabla \widetilde{\mathbf{V}}$ to finally maximize target items' ER@K:
\begin{equation}
  \label{eq_simple_opt}
  \begin{aligned}
    \mathop{argmax}\limits_{\{\nabla \widetilde{\mathbf{V}}^{t}, \nabla \widetilde{\mathbf{\Theta}}^{t}\}_{t=s}^{T-1}} ER@K(\mathbf{U}^{T},\mathbf{V}^{T},\mathbf{\Theta}^{T})\\
  \end{aligned}
\end{equation}
where $T$ is the total number of global epochs to train the FedRec and $s$ is the epoch that the attack starts at. 
$\nabla \widetilde{\mathbf{V}}^{t}$ and $\nabla \widetilde{\mathbf{\Theta}}^{t}$ are gradients generated by malicious users.
$\mathbf{U}^{T}$ is the learned embedding matrix of benign users, $\mathbf{V}^{T}$ and $\mathbf{\Theta}^{T}$ are the learned public parameters which depend on both benign users' and malicious users' aggregated gradients:
\begin{equation}
  \label{eq_add}
  \begin{aligned}
    &\mathbf{V}^{T} = \mathbf{V}^{0} - lr(\sum\limits_{t=0}^{T-1}\nabla \mathbf{V}^{t}  +\sum\limits_{t=s}^{T-1}\nabla \widetilde{\mathbf{V}}^{t})\\
    &\mathbf{\Theta}^{T} = \mathbf{\Theta}^{0} - lr(\sum\limits_{t=0}^{T-1}\nabla \mathbf{\Theta}^{t} + \sum\limits_{t=s}^{T-1}\nabla \widetilde{\mathbf{\Theta}}^{t})
  \end{aligned}
\end{equation}

\textbf{Attack Prior Knowledge.}
To make the threats of gradient poisoning attacks more realistic, the attack should utilize as less prior knowledge as possible.
In this paper, we assume the malicious users only know the public parameters $\mathbf{\Theta}^{t-1}$ and $\mathbf{V}^{t-1}$ received from the central server, which is consistent with the FedRec's protocol.

\textbf{Defense.}
The defense is launched at the central server side to defend against malicious users.
To work for most FedRecs, the defense should be able to seamlessly integrate into the basic FedRec protocol.
Besides, the defense is expected to eliminate poisoning attacks' influence with negligible side effects.

Algorithm~\ref{alg_fedrec} shows how to incorporate gradient poisoning attack and defense into a general federated recommendation framework.
\begin{algorithm}[!ht]
  \renewcommand{\algorithmicrequire}{\textbf{Input:}}
  \renewcommand{\algorithmicensure}{\textbf{Output:}}
  \caption{FedRec with poisoning attack and defense.} \label{alg_fedrec}
  \begin{algorithmic}[1]
    \Require global epoch $T$; local epoch $L$; learning rate $lr$, \dots
    \Ensure public parameter $\mathbf{V}$ and $\mathbf{\Theta}$, local client embedding $\mathbf{u}_{i}|_{i \in \mathcal{U}}$
    \State Initialize public parameter $\mathbf{V}^{0}$, $\mathbf{\Theta}^{0}$
    \For {each round t =1, ..., $T$}
      \State if $t$ is the epoch when the attack starts, $\mathcal{U} = \mathcal{U} \cup \widetilde{\mathcal{U}}$
      \State sample a fraction of clients $\mathcal{U}_{t-1}$ from $\mathcal{U}$
        \For{$u_{i}\in \mathcal{U}_{t-1}$ \textbf{in parallel}}
        \If{$u_{i}\in \widetilde{\mathcal{U}}$}
          \State // run on malicious client $u_{i}$
          \State execute attack algorithm (e.g. Algorithm~\ref{alg_attack})
        \Else
          \State // run on benign client $u_{i}$
          \State calculate $\nabla \mathbf{u}_{i}^{t-1}$, $\nabla \mathbf{V}_{i}^{t-1}$, $\nabla \mathbf{\Theta}_{i}^{t-1}$ using E.q.~\ref{eq_ori_loss}
          \State $\mathbf{u}_{i}^{t}\leftarrow$ update local private parameters using E.q.~\ref{eq_local_update}
          \State upload $\nabla \mathbf{V}_{i}^{t-1}$,  $\nabla \mathbf{\Theta}_{i}^{t-1}$ to the central server
        \EndIf
        \EndFor
      \State Run defense algorithm (e.g., Algorithm 3~\ref{alg_defense}) on the central server.
      \State $\mathbf{V}^{t}, \mathbf{\Theta}^{t}\leftarrow$ aggregate gradients using E.q.~\ref{eq_aggregate}
    \EndFor
    \end{algorithmic}
\end{algorithm}

\section{Our Poisoning Attack and Defense}
In this section, we present the details of our poisoning attack (PSMU) and defense (HiCS) methods.
They are also shown in Algorithms~\ref{alg_attack} and~\ref{alg_defense} with pseudo code respectively.

\subsection{PSMU: Poisoning with Synthetic Malicious Users}
The goal of PSMU is to promote target items $\widetilde{\mathcal{V}}$ maximally.
To achieve that, PSMU aims to maximize ER@K by uploading polluted gradients $\nabla \widetilde{\mathbf{\Theta}}$ and $\nabla \widetilde{\mathbf{V}}$ via malicious users, which is formulated in E.q.~\ref{eq_simple_opt}.
However, according to E.q.~\ref{eq_add}, $\mathbf{V}^{T}$ and $\mathbf{\Theta}^{T}$ depend on all previously uploaded gradients from both malicious users and benign users.
Since PSMU cannot access the benign clients' gradients $ \nabla \mathbf{V}^{t}$ and $ \nabla \mathbf{\Theta}^{t}$, it is infeasible to directly solve E.q.~\ref{eq_simple_opt}.
Therefore, instead of directly optimizing E.q.~\ref{eq_simple_opt}, we propose the following approximated optimization objective in E.q.~\ref{eq_iter_er} at each global epoch. The intuition is that if the malicious clients are generally consistent with benign clients, by greedily optimizing ER@K on malicious clients at each global epoch, the attack can  promote target items on benign clients~\cite{zhang2022pipattack}:
\begin{equation}
  \label{eq_iter_er}
  \begin{aligned}
    \mathop{argmax}\limits_{\{\nabla \widetilde{\mathbf{V}}^{t-1}, \nabla \widetilde{\mathbf{\Theta}}^{t-1}\}} ER@K(\mathbf{U}^{t-1},\mathbf{V}^{t-1} - lr\nabla \widetilde{\mathbf{V}}^{t-1},\mathbf{\Theta}^{t-1} - lr\nabla \widetilde{\mathbf{\Theta}}^{t-1})\\
  \end{aligned}
\end{equation}

However, E.q.~\ref{eq_iter_er} is still difficult to optimize by using gradient decent methods, since ER@K is discontinuous and non-differentiable (see E.q.~\ref{eq_er}).
Inspired by~\cite{rong2022fedrecattack}, we approximately optimize E.q.~\ref{eq_iter_er} by encouraging the target items' predicted preference scores to be higher than top-k recommended items', as follows:
\begin{equation}
  \label{eq_naive_loss}
  \mathcal{L}^{att} = \sum\limits_{u_{i}\in \mathcal{U}} \sum\limits_{v_{t} \in \widetilde{\mathcal{V}} \land v_{t} \notin \mathcal{V}_{i}^{+}} \sum\limits_{v_{j} \in \hat{\mathcal{V}_{i}} \land v_{j} \notin \widetilde{\mathcal{V}}} \sigma (\hat{r}_{ij} - \hat{r}_{it})
\end{equation}
To make the formula clear, we omit the time index in E.q.~\ref{eq_naive_loss}, since we greedily optimize it at every epoch.
To minimize E.q.~\ref{eq_naive_loss}, we need to know all benign users' embeddings $\mathbf{U}$ and their interacted item sets $\mathcal{V}_{i}^{+}$.
~\cite{rong2022fedrecattack} unrealistically assumes malicious users can obtain a proportion of interacted items from benign users and then aggregate these items' embeddings to approximate user embedding, which breaks FedRec's protocol.
To make the attack's threats realistic, the above information is not available in our attack settings.
~\cite{rong2022poisoning} simply uses vectors randomly sampled from a Gaussian distribution to represent user embeddings $\mathbf{U}$, which is not reasonable.

\begin{table}[!ht]
\centering
\caption{The Jaccard similarity of Top-10 popular items between randomly constructed users and real users when the model is converged.}\label{tb_proof}
\begin{tabular}{l|cccc}
\hline
                 & \multicolumn{2}{c}{\textbf{Fed-NCF}} & \multicolumn{2}{c}{\textbf{Fed-LightGCN}} \\ \hline
                 & \textbf{ML}       & \textbf{AZ}      & \textbf{ML}         & \textbf{AZ}         \\ \hline
\textbf{Jaccard} & 0.82              & 1.0              & 0.82                & 1.0                 \\ \hline
\end{tabular}
\end{table}

\textbf{Approximate with Synthetic Users.}
The goal of E.q.~\ref{eq_naive_loss} is to enforce target items' prediction scores to be higher than top-k recommended items, so as to improve the target items'  exposure rate.
Since we cannot access benign users' embeddings, we propose to construct the synthetic users with randomly selected items.
Intuitively, if the target items can appear in the top-K recommendation lists of randomly constructed users, the target items could also be promoted to real users by the recommender model.
In other words, the popular items in real users and synthetic users are consistent.
Table~\ref{tb_proof} provides a proof-of-concept.
We measure the similarity of top-10 popular items in randomly constructed users' and real users' recommendation lists with Jaccard similarity. The results show that popular items between these two kinds of users are highly similar.
As a result, we can promote a target item to real users by making it popular with our synthetic users.
To achieve that, when a malicious user $\widetilde{u}_{i}$ participates in FedRec training process, it randomly selects $\alpha$ items (except for target items) as its interacted items $\widetilde{\mathcal{V}}_{i}^{+}$ and constructs the training set $\widetilde{\mathcal{D}}_{i}$.
Note that at different epochs, the malicious user $\widetilde{u}_{i}$ has different $\widetilde{\mathcal{V}}_{i}^{+}$ and training set $\widetilde{\mathcal{D}}_{i}$, so that even with a small set of malicious users, we can simulate many synthetic users.
As a result, E.q.~\ref{eq_naive_loss} is transformed to the following loss function:
\begin{equation}
  \label{eq_avid_loss}
  \widetilde{\mathcal{L}}^{att} = \sum\limits_{\widetilde{u}_{i}\in \widetilde{\mathcal{U}}} \sum\limits_{v_{t} \in \widetilde{\mathcal{V}} \land v_{t} \notin \widetilde{\mathcal{V}}^{+}_{i}} \sum\limits_{v_{j} \in \hat{\widetilde{\mathcal{V}}_{i}} \land v_{j} \notin \widetilde{\mathcal{V}}} \sigma (\hat{r}_{ij} - \hat{r}_{it})
\end{equation}
where $\hat{\widetilde{\mathcal{V}}_{i}}$ are the set of items that have the highest prediction scores for malicious user $\widetilde{u}_{i}$.

\textbf{Compete with Alternative Products.} 
By optimizing E.q.~\ref{eq_avid_loss}, the target items are competitive with the recommended items.
To further improve the target items' competition, we further attempt to improve target items' prediction scores by adding the alternative products of the target items to enlarge the competition item set, since alternative products are interchangeable and competitive with the target items.
As the attack cannot use any prior knowledge, malicious user $\widetilde{u}_{i}$ selects items that have higher item embedding similarity with target items, meanwhile, have relatively higher preference scores as alternative products $\widetilde{\mathcal{V}}_{i}^{ap}$.
Therefore, E.q.~\ref{eq_avid_loss} is further transformed to:
\begin{equation}
  \label{eq_final_loss}
  \widetilde{\mathcal{L}}_{ap}^{att} = \sum\limits_{\widetilde{u}_{i}\in \widetilde{\mathcal{U}}} \sum\limits_{v_{t} \in \widetilde{\mathcal{V}} \land v_{t} \notin \widetilde{\mathcal{V}}^{+}_{i}} \sum\limits_{v_{j} \in \{\hat{\widetilde{\mathcal{V}}_{i}} \cup \widetilde{\mathcal{V}}_{i}^{ap}\}  \land v_{j} \notin \widetilde{\mathcal{V}}} \sigma (\hat{r}_{ij} - \hat{r}_{it})
\end{equation}

\textbf{Calculate Poisoned Gradients.} 
The attack objective $\widetilde{\mathcal{L}}^{att}$ now only relies on $\widetilde{\mathbf{U}}$, $\mathbf{V}$, $\mathbf{\Theta}$, $\hat{\widetilde{\mathcal{V}}_{i}}$, and $\widetilde{\mathcal{V}}_{i}^{ap}$, i.e. $\widetilde{\mathcal{L}}^{att}(\widetilde{\mathbf{U}},\mathbf{V},\mathbf{\Theta},\hat{\widetilde{\mathcal{V}}_{i}}, \widetilde{\mathcal{V}}_{i}^{ap})$.
$\mathbf{V}$ and $\mathbf{\Theta}$ can be directly obtained from the central server.
$\widetilde{\mathcal{V}}_{i}^{ap}$ can be created based on $\mathbf{V}$.
$\hat{\widetilde{\mathcal{V}}_{i}}$ can be calculated based on $\mathbf{V}$ and $\widetilde{\mathbf{U}}$.
Therefore, to optimize $\widetilde{\mathcal{L}}^{att}$, we first need to calculate $\widetilde{\mathbf{U}}$.

At epoch $t$, to calculate $\widetilde{\mathbf{U}}^{t-1}$, the malicious users first randomly initialize their corresponding user embeddings.
Then, they fix the received public parameters and only update user embeddings to optimize local recommendation loss on their synthetically constructed datasets $\widetilde{\mathcal{D}}^{t-1}$:

\begin{equation}
  \label{eq_avid_user}
  \widetilde{\mathbf{U}}^{t-1} = \mathop{argmin}\limits_{\widetilde{\mathbf{U}}^{t-1}} \mathcal{L}^{rec}(\widetilde{\mathbf{U}}^{t-1}, \mathbf{V}^{t-1}, \mathbf{\Theta}^{t-1}, \widetilde{\mathcal{D}}^{t-1})
\end{equation}

Then, for each malicious user, we use $\widetilde{\mathbf{u}}_{i}^{t-1}$, $\mathbf{V}^{t-1}$ and $\mathbf{\Theta}^{t-1}$ to get items $\hat{\widetilde{\mathcal{V}}_{i}}$ and $\widetilde{\mathcal{V}}_{i}^{ap}$.
Finally, we fix the malicious user embeddings $\widetilde{\mathbf{U}}^{t-1}$ and finetune $\mathbf{V}^{t-1}$ and $\mathbf{\Theta}^{t-1}$ to minimize $\widetilde{\mathcal{L}}^{att}$ as follows:
\begin{equation}
  \label{eq_poison_grad}
  \begin{aligned}
    &\nabla \widetilde{\mathbf{V}}^{t-1} = \frac{\partial}{\partial \mathbf{V}^{t-1}} \widetilde{\mathcal{L}}_{ap}^{att}(\widetilde{\mathbf{U}}^{t-1},\mathbf{V}^{t-1},\mathbf{\Theta}^{t-1},\hat{\widetilde{\mathcal{V}}}^{t-1}_{i}, \widetilde{\mathcal{V}}^{ap,t-1}_{i})  \\
    &\nabla \widetilde{\mathbf{\Theta}}^{t-1} = \frac{\partial}{\partial \mathbf{\Theta}^{t-1}} \widetilde{\mathcal{L}}_{ap}^{att}(\widetilde{\mathbf{U}}^{t-1},\mathbf{V}^{t-1},\mathbf{\Theta}^{t-1},\hat{\widetilde{\mathcal{V}}}^{t-1}_{i}, \widetilde{\mathcal{V}}^{ap,t-1}_{i})  \\
  \end{aligned}
\end{equation}

Furthermore, to avoid significant side effects on FedRec's performance, for $\nabla \widetilde{\mathbf{V}}^{t-1}$, we only upload the poisoned gradients of the target items.
\begin{equation}
  \label{eq_refine}
  \nabla \widetilde{\mathbf{V}}^{t-1}=
  \begin{cases}
  \mathbf{0}& v_{m} \notin \widetilde{V}\\
  \nabla \widetilde{\mathbf{V}}^{t-1}_{m}& v_{m} \in \widetilde{V}
  \end{cases} \quad m=0,1,\dots,\left| \mathcal{V} \right|
\end{equation}
Algorithm~\ref{alg_attack} describes PSMU with pseudo code.
\begin{algorithm}[!ht]
  \renewcommand{\algorithmicrequire}{\textbf{Input:}}
  \renewcommand{\algorithmicensure}{\textbf{Output:}}
  \caption{PSMU: Poisoning with Synthetic Malicious Users} \label{alg_attack}
  \begin{algorithmic}[1]
     
    \Require public parameters $\mathbf{V}^{t-1}$, $\mathbf{\Theta}^{t-1}$
    \Ensure public parameter poisoned gradients $\nabla \widetilde{\mathbf{V}}_{i}^{t-1}$, $\nabla \widetilde{\mathbf{\Theta}}_{i}^{t-1}$
    \State // run on malicious client $\widetilde{u}_{i}$
    \State randomly construct training set $\widetilde{\mathcal{D}}^{t-1}_{i}$
    \State calculate synthetic user embedding $\widetilde{\mathbf{u}}_{i}^{t-1}$ using E.q.~\ref{eq_avid_user}
    \State calculate $\nabla \widetilde{\mathbf{V}}_{i}^{t-1}$, $\nabla \widetilde{\mathbf{\Theta}}_{i}^{t-1}$ using E.q.~\ref{eq_poison_grad}
    \State $\nabla \widetilde{\mathbf{V}}_{i}^{t-1}\leftarrow$ constraint $\nabla \widetilde{\mathbf{V}}_{i}^{t-1}$ using E.q.~\ref{eq_refine}
    \State upload $\nabla \widetilde{\mathbf{V}}_{i}^{t-1}$,  $\nabla \widetilde{\mathbf{\Theta}}_{i}^{t-1}$ to the central server
    \end{algorithmic}
\end{algorithm}

\subsection{HiCS: Hierarchical Gradient Clipping and
Sparsification Update for Defense}\label{sec_hics}
The effectiveness of PSMU reveals the vulnerability of FedRecs, but there is not any effective defense solution against such kind of poisoned gradient attacks on FedRecs as mentioned in Section~\ref{sec_intro}.
To fill this gap, we take the first step to propose an effective defense method against poisoning attacks in FedRec, HiCS, which is based on clipping and sparsified updating.
Note that HiCS only processes the gradients of item embeddings, since a large number of benign users can counteract the poisoned gradients of the parameters $\mathbf{\Theta}$.

\textbf{Gradient Clipping.} %
To alleviate poisoned gradients' effects, one naive way is to clip all uploaded gradients, so that each malicious user can only contribute at most $\rho$ with $\ell_{p}$ normalization (we use $\ell_{2}$ in this paper).
\begin{equation}\label{eq_easy_clip}
  \nabla \mathbf{V}_{i}^{t-1} = \nabla \mathbf{V}_{i}^{t-1} \cdot \min \left(1, \frac{\rho}{\left\| \nabla \mathbf{V}_{i}^{t-1} \right\|_{p}}\right)
\end{equation}
where $\nabla \mathbf{V}_{i}^{t-1}$ is the gradients uploaded by user $u_{i}$ who can be either benign or malicious.
After applying gradient clipping, the mass of poisoned gradients is constrained:
\begin{equation}
  \label{eq_constrain}
    \left\| \nabla \widetilde{\mathbf{V}}^{t-1}\right\|_{p} = \left\| \sum\limits_{\widetilde{u}_{i}\in \widetilde{\mathcal{U}}_{t-1}} \nabla \widetilde{\mathbf{V}}^{t-1}_{i}\right\|_{p} \leq \rho \left| \widetilde{\mathcal{U}}_{t-1} \right|
\end{equation}
However, as shown in E.q.~\ref{eq_constrain}, the attack can be still effective by compromising more malicious users, i.e., increasing $\left| \widetilde{\mathcal{U}}_{t-1} \right|$.
As a result, simply using gradient clipping cannot significantly reduce the attacker's effectiveness when increasing malicious user numbers.

\textbf{Clipping with Sparsified Updating.} 
To further improve FedRec's robustness against poisoning attacks, we conduct another gradient clipping combined with sparsified updating.
After clipping the gradients $\nabla\mathbf{V}_{i}^{t-1}$, the server aggregates these gradients and stores them in a memory bank $\mathbf{W}^{t-1}$.
$\mathbf{W}^{t-1}$ is a matrix with $\left| \mathcal{V} \right|$ rows.
The server selects the top $\gamma$ item embedding gradients with the largest magnitudes from $\mathbf{W}^{t-1}$, and zeros out these gradients from $\mathbf{W}^{t-1}$.
The top $\gamma$ gradients $\nabla_{top}\mathbf{V}^{t-1}$ will be used to update the global item embedding table.
As $\left| \widetilde{\mathcal{U}}_{t-1} \right|$ is limited, the magnitude of poisoned gradients $\nabla \widetilde{\mathbf{V}}^{t-1}$ would not be big enough at the early stage of the attack so that these poisoned gradients would be less likely to  be selected as top $\gamma$, hence delaying the attack.
These poisoned gradients need to wait until the target items' accumulative gradient magnitudes are large enough.
However, during the accumulation, benign users' gradients would have increasingly higher chances to dilute these poisoned gradients.

Even though, relying on benign users' gradients to neutralize poisoned gradients is unstable and may be less effective when fewer benign users interact with the target items (i.e., unpopular items).
Therefore, we further apply a gradient clipping with adaptive gradient limits on $\nabla_{top}\mathbf{V}^{t-1}$.
Specifically, we utilize the average normalization of $\nabla_{top}\mathbf{V}_{i}^{t-1}$ as the gradient limits and then clip all the gradients:
\begin{equation}\label{eq_adaptive_clip}
  \nabla_{top} \mathbf{V}_{i}^{t-1} = \nabla_{top} \mathbf{V}_{i}^{t-1} \cdot \min \left(1, \frac{{\rm avg}(\left\| \nabla_{top} \mathbf{V}_{i}^{t-1} \right\|_{p})}{\left\| \nabla_{top} \mathbf{V}_{i}^{t-1} \right\|_{p}}\right)
\end{equation}
As a result, even when $\left| \widetilde{\mathcal{U}}_{t-1} \right|$ is increased, the poisoned gradients still have limited mass since we clip each item embedding's accumulative gradients with average normalization of $\nabla_{top}\mathbf{V}^{t-1}$.
Algorithm~\ref{alg_defense} shows how HiCS works with pseudo code.

\begin{algorithm}[!ht]
  \renewcommand{\algorithmicrequire}{\textbf{Input:}}
  \renewcommand{\algorithmicensure}{\textbf{Output:}}
  \caption{HiCS:  Hierarchical Gradient Clipping and Sparsification Update for Defense} \label{alg_defense}
  \begin{algorithmic}[1]
    \Require uploaded item embedding gradients $ \nabla \mathbf{V}^{t-1}$
    \Ensure processed gradients $\nabla_{top}\mathbf{V}^{t-1}$
    \State initialize memory bank $\mathbf{W}^{t-1}$ to $\mathbf{0}$ if $t=1$
    \State $ \nabla \mathbf{V}^{t-1}\leftarrow$ clip gradient $ \nabla \mathbf{V}^{t-1}$ using E.q.~\ref{eq_easy_clip}
    \State store $ \nabla \mathbf{V}^{t-1}\leftarrow$ in memory bank: $\mathbf{W}^{t-1} = \mathbf{W}^{t-1} + \nabla \mathbf{V}^{t-1}$
    \State extract top $\gamma$ gradients: $\nabla_{top}\mathbf{V}^{t-1}={\rm top} (\mathbf{W}^{t-1}, \gamma)$
    \State zero out extracted gradients: $\mathbf{W}^{t} = \mathbf{W}^{t-1} - \nabla_{top}\mathbf{V}^{t-1}$
    \State $\nabla_{top}\mathbf{V}^{t-1}\leftarrow$ clip with adaptive gradient limits using E.q.~\ref{eq_adaptive_clip}
    \end{algorithmic}
\end{algorithm}


\section{Experiments}
In this section, we conduct extensive experiments to explore the following research questions (RQs):
\begin{itemize}
  \item \textbf{RQ1}. How is the effectiveness of our attack (PSMU) compared with gradient poisoning attack baselines?
  \item \textbf{RQ2}. Can the proposed defense (HiCS) effectively defend against gradient poisoning attacks?
  \item \textbf{RQ3}. How is the effectiveness of our defense (HiCS) compared with defense baselines?
  \item \textbf{RQ4}. The impact analysis of malicious user proportion for both proposed attack and defense.
\end{itemize}

\subsection{Datasets}
In this paper, we adopt two popular federated recommendation datasets for evaluation, namely MovieLens-1M (ML)~\cite{harper2015movielens} and Amazon Digital Music (AZ)~\cite{mcauley2015image}.
ML contains $1,000,208$ ratings involving $6,040$ users and $3,706$ movies. 
AZ includes $169,781$ interactions with $16,566$ users and $11,797$ products, and all users have at least $5$ interactions with different products.
Following~\cite{rong2022poisoning,zhang2022pipattack}, we binarize the user-item ratings, where all ratings are transformed to $r_{ij}=1$ and negative instances are sampled with $1:4$ ratio.
$80\%$ and $20\%$ data are divided into training and test set.

\subsection{Evaluation Protocol}
Following~\cite{bhagoji2019analyzing,zhang2022pipattack}, we first train FedRecs without attack for several epochs.
Then, at a certain epoch, an attack method is launched.
The FedRec is trained until it is converged.
We select the most unpopular items as target items.
When evaluating defense methods, the tested defense method will be activated once the FedRec's training starts, as the system cannot predict when the attack will be launched.

The evaluation of attack methods is from two aspects: (1) whether the attack method can increase the average exposure rate (ER@5) of target items; (2) whether the attack significantly degenerates the FedRec's recommendation performance (HR@20).
An ideal targeted attack method should improve the exposure rate meanwhile cause fewer side effects on FedRec's performance.
The evaluation of defense methods is similar to the attack but their goal is to prevent attacks from manipulating target items' exposure rate meanwhile cause fewer side effects to the FedRec.

\begin{figure*}[!htbp]
  \centering
  \subfloat[The trend of exposure rate of target items with $0.1\%$ malicious users.]{\includegraphics[width=1.\textwidth]{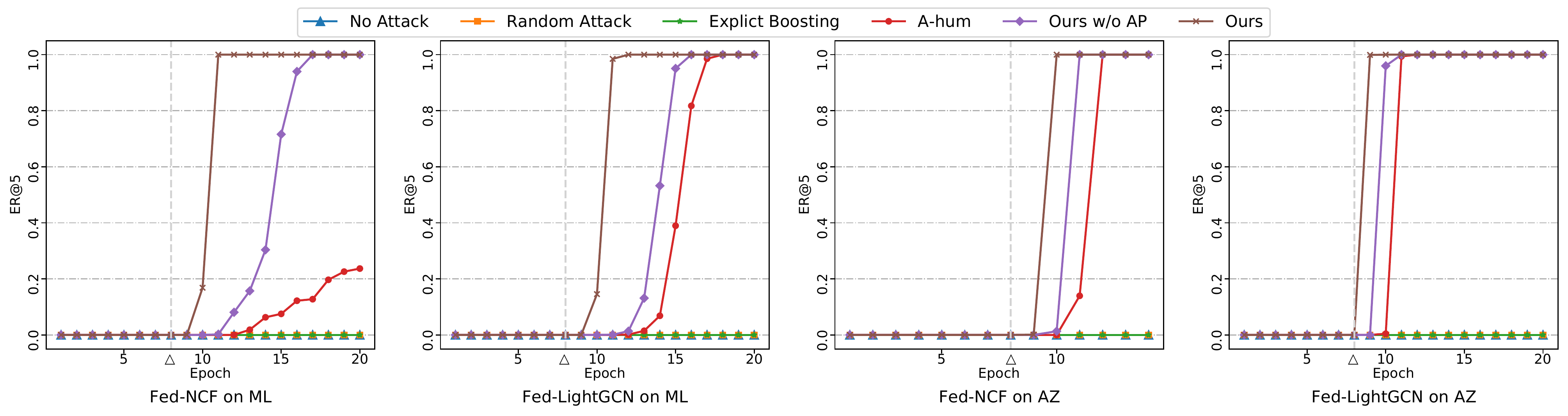}\label{fig_att_er_base}}
  \hfil
  \subfloat[The trend of FedRecs' recommendation performance. With fewer ($0.1\%$) malicious users, all attack methods cause negligible side effects to FedRecs.]{\includegraphics[width=1.\textwidth]{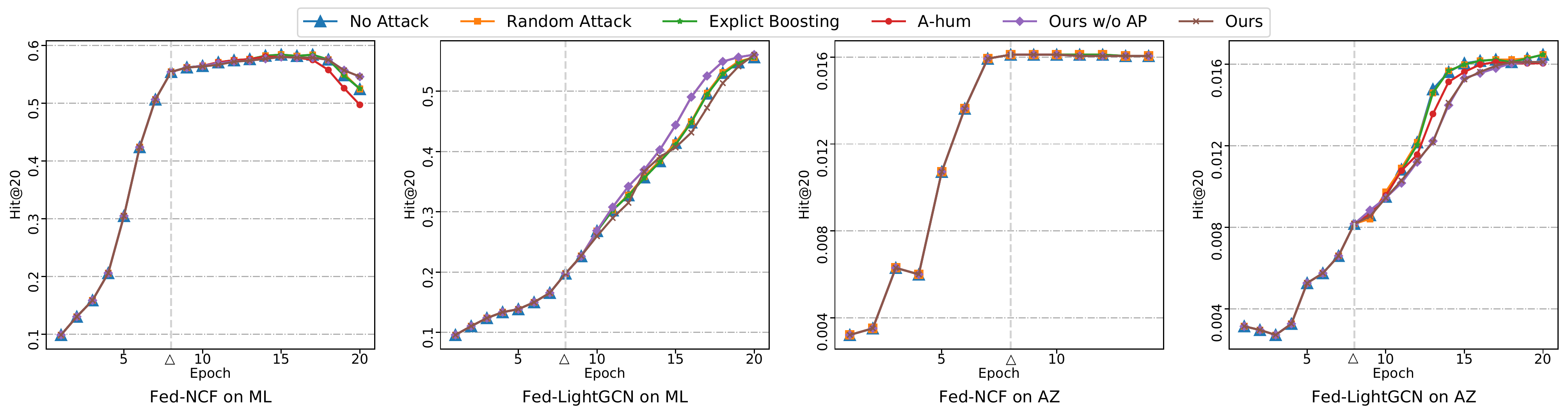}\label{fig_att_hit_base}}
  \caption{The comparison of attack performance between PSMU and attack baselines, the attack starts at $\triangle$.}\label{fig_att_base}
\end{figure*}

\begin{figure*}[!htbp]
  \centering
  \includegraphics[width=1.\textwidth]{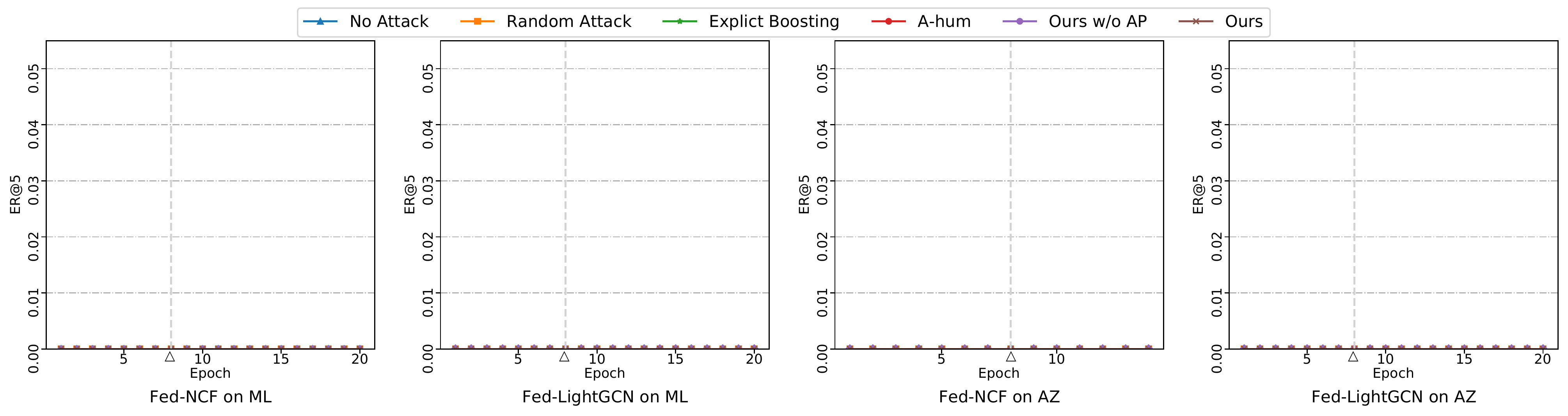}
  \caption{Attacks' performance for FedRecs equipped with HiCS defense.}\label{fig_hics_against_att_baselines}
\end{figure*}

\subsection{Baselines}

\textbf{Attack Baselines.}
We consider both data poisoning attacks and gradient poisoning attacks as our baselines.
However, some data poisoning attacks (e.g. Bandwagon attacks~\cite{gunes2014shilling}) and gradient poisoning attacks (e.g. FedRecAttack~\cite{rong2022fedrecattack} and PipAttack~\cite{zhang2022pipattack}) rely on prior knowledge.
As we focus on the attack setting without prior knowledge, these attack methods are not adopted as baselines for a fair comparison.
Therefore, we choose the following baselines:
\begin{itemize}
  \item \textbf{No Attack}. This method shows the original exposure rate of the target items and the FedRecs' normal performance.
  \item \textbf{Random Attack (RA)}~\cite{kapoor2017review}. It is a simple data poisoning attack that injects malicious users with both random interactions and target item interactions.
  \item \textbf{Explicit Boosting (EB)}. It is a component of PipAttack~\cite{zhang2022pipattack} which does not rely on prior knowledge.
  \item \textbf{A-hum}~\cite{rong2022poisoning}. The current state-of-the-art gradient poisoning attack without relying on prior knowledge in FedRecs. 
  \item \textbf{Ours w/o AP}. It is PSMU that removes alternative products, i.e., using E.q.~\ref{eq_avid_loss} as the attack's optimization objective.
\end{itemize}

\textbf{Defense Baselines.} As mentioned in Section~\ref{sec_hics}, none of the existing defense methods is specifically proposed for FedRecs and most defense methods in federated learning cannot be directly applied to FedRecs. 
For the purpose of comparison, we choose the following defense methods which are popular in federated learning and are still applicable to our FedRec setting as defense baselines.
\begin{itemize}
  \item \textbf{No Defense}. This method shows the original FedRec's performance under certain attacks.
  \item \textbf{Item-level Krum}. The original Krum~\cite{blanchard2017machine} cannot be applied to FedRec, since the uploaded gradients from different clients are not comparable. To make Krum applicable in our problem setting, we propose Item-level Krum. For each item,  it selects the embedding gradient that is closest to the mean of all the other clients'  uploaded gradients of the item as the aggregated gradient.
  \item \textbf{Median}~\cite{yin2018byzantine}. It sorts the values of uploaded parameters and chooses the median as the aggregated value. 
  \item \textbf{Trimmed Mean}~\cite{yin2018byzantine}. It aggregates gradients by removing the largest and smallest value of a parameter and calculating the mean of the remaining ones.
  \item \textbf{$\ell_{2}$ clipping}~\cite{guerraoui2018hidden}. It is a component of our HiCS, that clips all gradients using $\ell_{2}$ normalization before aggregation.
\end{itemize}

\begin{figure*}[!htbp]
  \centering
  \subfloat[The comparison of HiCS and defense baselines against PSMU. Our defense reduces target items' exposure rate in all cases.]{\includegraphics[width=1.\textwidth]{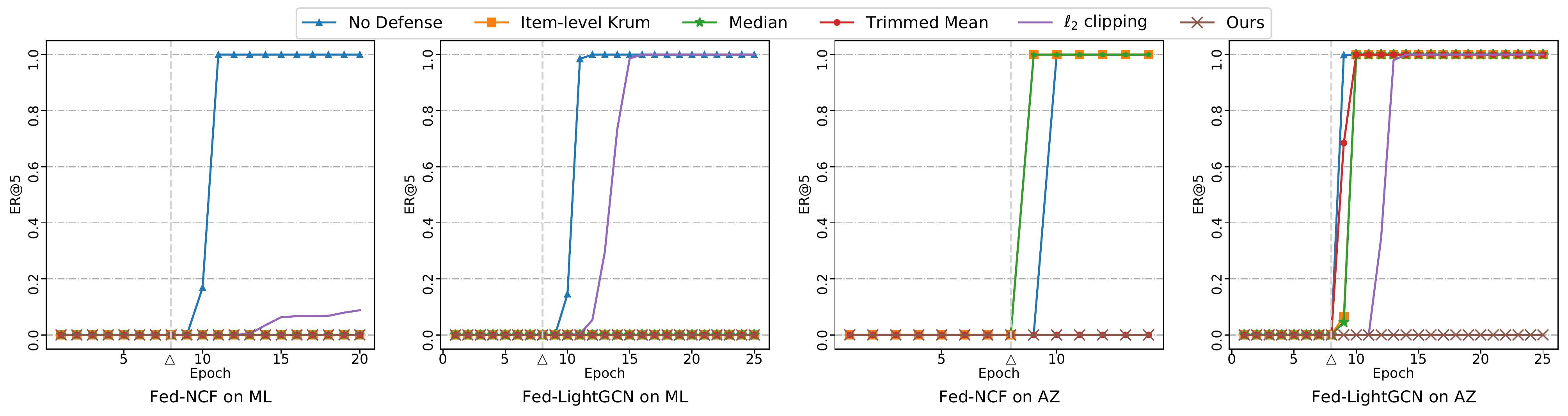}\label{fig_hics_against_def_baselines}}
  \hfil
  \subfloat[The effects of different defense methods on FedRecs' recommendation performance. Our defense even improves FedRecs' performance on AZ since it provides regularization to alleviate over-fitting on the sparse dataset. ]{\includegraphics[width=1.\textwidth]{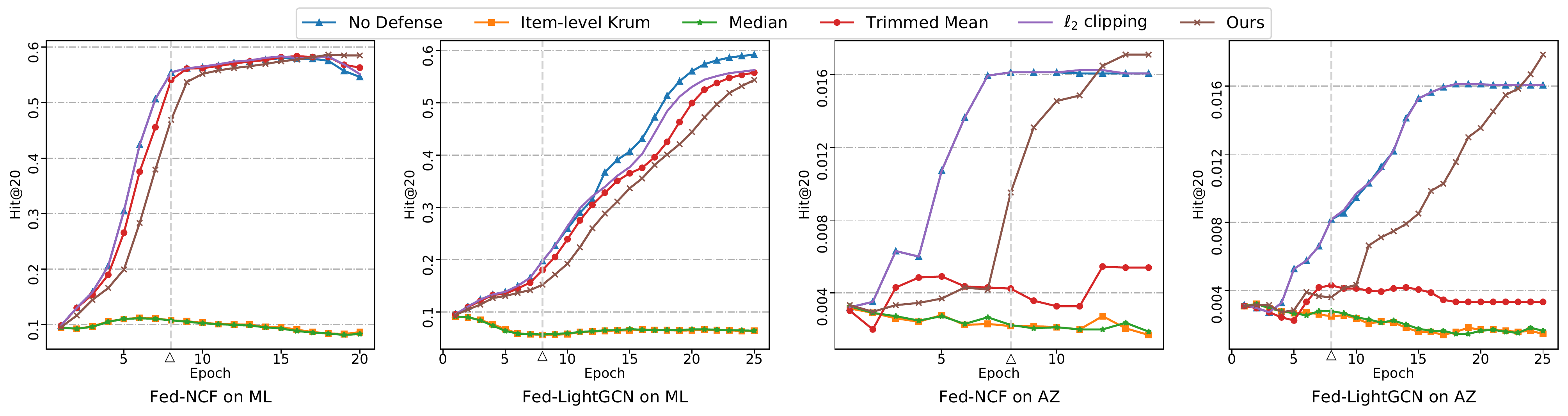}\label{fig_hics_against_def_baselines_hit}}
  \caption{The comparison of all defenses.}\label{fig_def_base}
\end{figure*}
\begin{table}[!ht]
  \centering
  \caption{Summarization of Fig.~\ref{fig_def_base}. In a cell, left \checkmark or $\times$ represents whether the defense can reduce target items' exposure rate (Fig.~\ref{fig_hics_against_def_baselines}), right \checkmark or $\times$ represents whether the FedRec's performance is not influenced (Fig.~\ref{fig_hics_against_def_baselines_hit}). Both values are \checkmark representing that the defense is effective.}\label{tb_def_summary}
  \begin{tabular}{l|cccc}
  \hline
  \textbf{Defense}         & \multicolumn{2}{c}{\textbf{Fed-NCF}} & \multicolumn{2}{c}{\textbf{Fed-LightGCN}} \\ \hline
  \textbf{}                & \textbf{ML}       & \textbf{AZ}      & \textbf{ML}         & \textbf{AZ}         \\ \hline
  \textbf{No Defense}      & $\times$/\checkmark               & $\times$/\checkmark              & $\times$/\checkmark                 & $\times$/\checkmark                 \\
  \textbf{Item-level Krum} & \checkmark/$\times$               & $\times$/$\times$              & \checkmark/$\times$                 & $\times$/$\times$                 \\
  \textbf{Median}          & \checkmark/$\times$               & $\times$/$\times$              & \checkmark/$\times$                 & $\times$/$\times$                 \\
  \textbf{Trimmed Mean}    & \checkmark/\checkmark               & \checkmark/$\times$              & \checkmark/\checkmark                 & $\times$/$\times$                 \\
  \textbf{$\ell_{2}$ clipping}     & \checkmark/\checkmark               & \checkmark/\checkmark              & $\times$/\checkmark                 & $\times$/\checkmark                 \\
  \textbf{Ours}            & \checkmark/\checkmark               & \checkmark/\checkmark              & \checkmark/\checkmark                 & \checkmark/\checkmark                 \\ \hline
  \end{tabular}
  \end{table}
\subsection{Parameter Settings}
For both Fed-NCF and Fed-LightGCN, the dimension of user and item embeddings is $32$.
$3$ feedforward layers with dimensions $64$, $32$, and $16$ are used to process the concatenated user and item embeddings.
The layer of LightGCN propagation is $1$.
Adam~\cite{kingma2014adam} with learning rate $lr=0.001$ is adopted.
The attack starts at $8$th global epoch.
$\alpha, \gamma, \rho$ are $30$, $10\%$, and $1.0$, respectively.
The proportion of malicious users $\xi$ is set to $0.1\%$ without specific mention.
In Section~\ref{sec_xi}, we investigate smaller $\xi$ to show PSMU's performance with extremely fewer malicious users.

\subsection{Effectiveness of PSMU (RQ1)}
To show the superiority of PSMU, we compare its attack performance and its effects on FedRecs' performance with all attack baselines in Fig.~\ref{fig_att_base}.
In Fig.~\ref{fig_att_er_base}, our PSMU outperforms all baselines on all datasets with all FedRecs.
Specifically, when attacking Fed-NCF on ML, A-hum can only achieve about $0.23$ ER@5 scores, while PSMU exposes the target item to all users.
In the other three cases, our attack and A-hum both achieve $1.0$ scores, but PSMU uses fewer epochs.
Besides, the comparison between our attack and Ours w/o AP indicates that alternative products can accelerate the item promotion process, since PSMU takes fewer epochs to reach $1.0$ ER@5 scores.
In fact, in all cases, PSMU promotes target items to all clients within at most $3$ epochs, which is much faster than all other baselines.
Random Attack and Explicit Boosting cannot work with only $0.1\%$ malicious users.

Fig.~\ref{fig_att_hit_base} illustrates the side effects of attacks on FedRecs' recommendation performance.
To ensure the stealthiness and effectiveness of item promotion, all attacks attempt to avoid significant side effects on recommendation performance.
As shown in Fig.~\ref{fig_att_hit_base}, all attack methods produce fewer side effects on recommendation performance.
This is because the number of malicious users is limited, i.e., there are only $0.1\%$ malicious users in the training process.

\subsection{HiCS against poisoning attacks (RQ2)}
The effectiveness of poisoning attacks reveals the security hole of vanilla FedRecs, however, no previous work explores defense methods against these attacks.
Therefore, in this paper, we propose HiCS to fix the security hole of vanilla FedRecs.
Specifically, we incorporate HiCS in FedRec from the initial stage since the server cannot predict when the attack will occur.
Fig.~\ref{fig_hics_against_att_baselines} shows the results of attack performance against FedRecs equipped with HiCS.
As we can see, all attacks obtain $0.0$ ER@5 values, which indicates that HiCS successfully defend against these attacks.
The side effects of HiCS on the recommender system are presented and analyzed in Section~\ref{sec_rq3} to avoid repetition.

\subsection{Comparison of HiCS and defense baselines (RQ3)}\label{sec_rq3}
In Section~\ref{sec_rq3}, we show that HiCS successfully defends against all selected attack methods. Here, we use our PSMU as the attack baseline and compare HiCS with defense baselines to show the superiority of our proposed defense method in Fig.~\ref{fig_def_base}.
As mentioned before, an effective defense method should satisfy two requirements: (1) compromise the attack's performance, i.e., reduce target items' exposure rate to normal values; (2) cause fewer side effects on recommendation performance.
Fig.~\ref{fig_hics_against_def_baselines} and Fig.~\ref{fig_hics_against_def_baselines_hit} show the evaluation of these two aspects respectively.
For convenient comparison, we summarize Fig.~\ref{fig_def_base} in Table.~\ref{tb_def_summary}.
From Table.~\ref{tb_def_summary}, we can know that our defense method keeps effective in all cases, since it satisfies the above two requirements.
Item-level Krum and Median are not effective in all cases since they destroy FedRecs' recommendation performance.
Trimmed Mean is ineffective on AZ dataset.
$\ell_{2}$ clipping does not compromise FedRecs' performance, however, it is too weak to defend against PSMU in Fed-LightGCN.
To sum up, only our defense methods keep consistent effectiveness in all cases.

\begin{figure}[!h]
  \centering
  \subfloat[PSMU with different number of malicious users on ML.]{\includegraphics[width=0.45\textwidth]{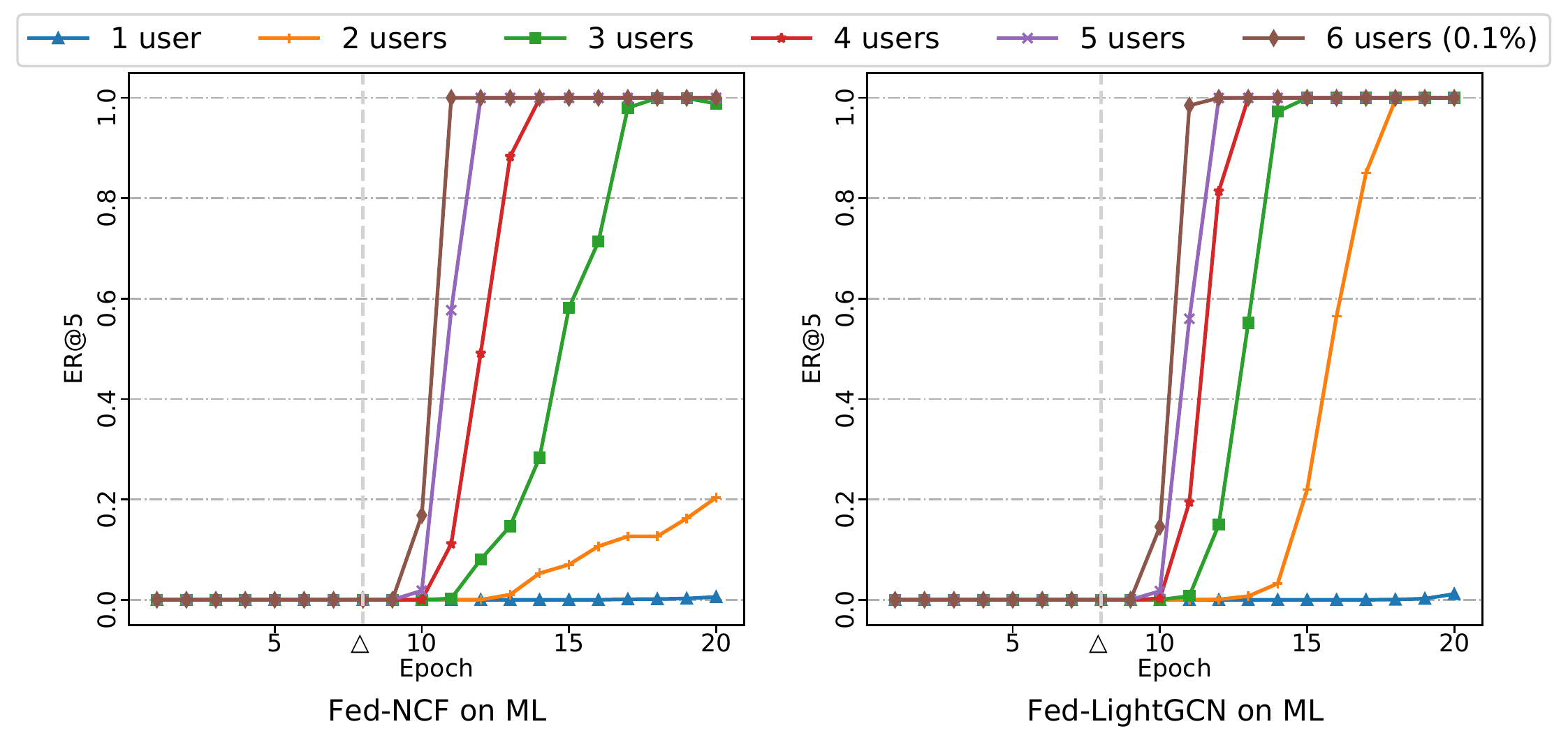}\label{fig_att_proportion_ml}}
  \hfil
  \subfloat[PSMU with different number of malicious users on AZ.]{\includegraphics[width=0.42\textwidth]{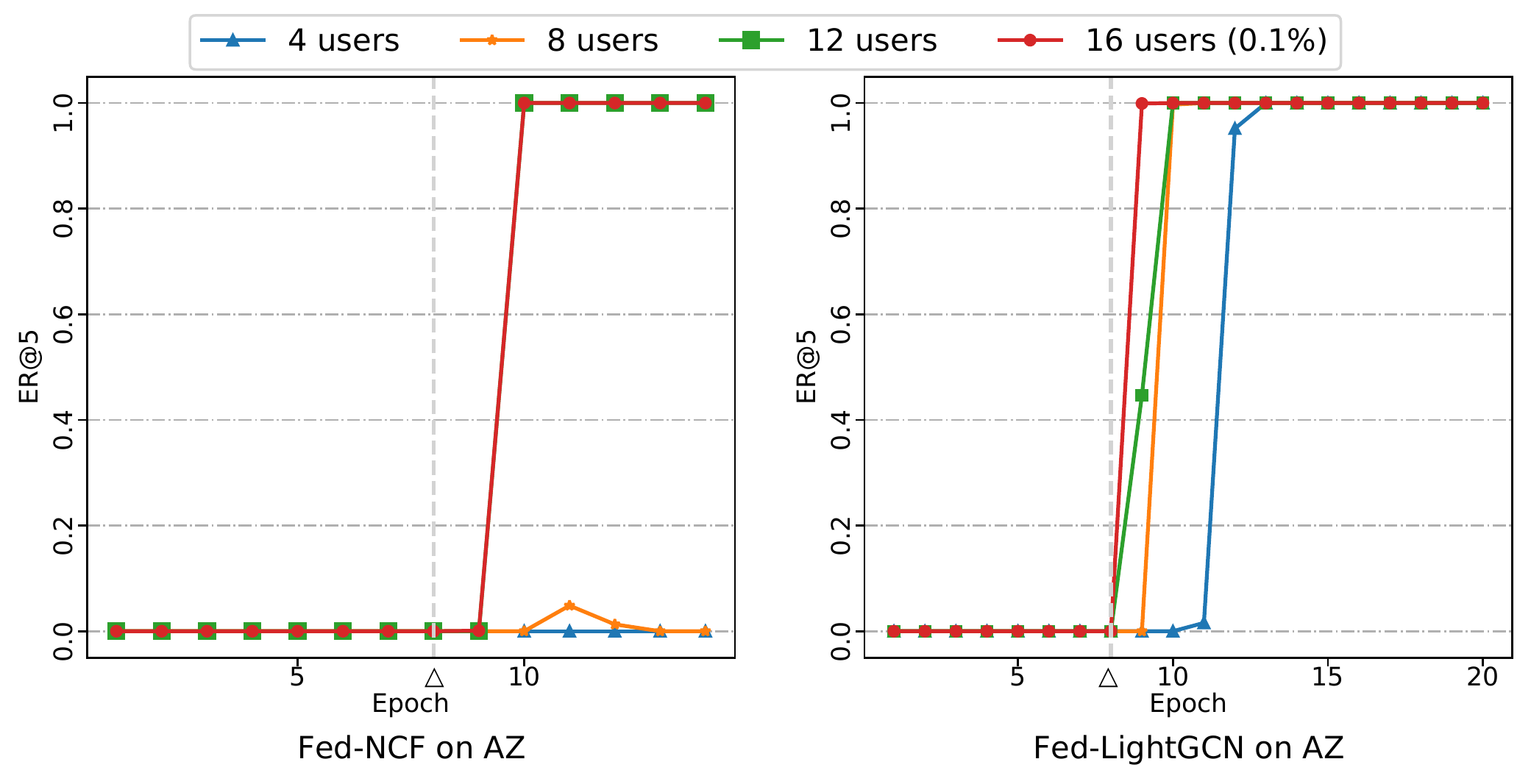}\label{fig_att_proportion_az}}
  \caption{The impact of malicious user number on PSMU.}\label{fig_att_proportion}
  \vspace{-8pt}
\end{figure}

  \begin{table}[!htbp]
    \centering
    \setlength\tabcolsep{2.pt}
    \caption{The impacts of malicious user proportion $\xi$ on HiCS. In a cell, left/right value is from ML/AZ. ``su" shorts for sparsified updating.}\label{tb_defense_proportion}
    \resizebox{0.47\textwidth}{!}{
    \begin{tabular}{l|ccc|ccc}
    \hline
    \textbf{defense method}                                                 & \multicolumn{3}{c|}{\textbf{Fed-NCF}}                              & \multicolumn{3}{c}{\textbf{Fed-LightGCN}}                 \\ \hline
     (ML, AZ)                                                        & $\xi$=0.1\%        & $\xi$=1.0\%            & $\xi$=10\%                   & $\xi$=0.1\%        & $\xi$=1.0\%        & $\xi$=10\%               \\ \hline
    \textbf{no defense}                                               & $\left(1.0,1.0\right)$          & $\left(1.0,1.0\right)$              & $\left(1.0,1.0\right)$                       & $\left(1.0,1.0\right)$          & $\left(1.0,1.0\right)$          & $\left(1.0,1.0\right)$                   \\
    \textbf{$\ell_{2}$ clipping}                                              & $\left(0.1,\mathbf{0.0}\right)$      & $\left(1.0,1.0\right)$              & $\left(1.0,1.0\right)$                      & $\left(1.0,1.0\right)$          & $\left(1.0,1.0\right)$           & $\left(1.0,1.0\right)$                     \\
    \textbf{$\ell_{2}$ clipping + su}                                              & $\left(\mathbf{0.0},\mathbf{0.0}\right)$      & $\left(1.0,1.0\right)$              & $\left(1.0,1.0\right)$                      & $\left(\mathbf{0.0},\mathbf{0.0}\right)$          & $\left(1.0,1.0\right)$           & $\left(1.0,1.0\right)$                     \\
    \textbf{ours} & $\left(\mathbf{0.0},\mathbf{0.0}\right)$ & $\left(\mathbf{0.0},\mathbf{0.0}\right)$ & $\left(\mathbf{0.0},\mathbf{0.0}\right)$  & $\left(\mathbf{0.0},\mathbf{0.0}\right)$ & $\left(\mathbf{0.0},\mathbf{0.0}\right)$ & $\left(\mathbf{0.0},\mathbf{0.0}\right)$  \\ \hline
    \end{tabular}}
    \end{table}

\subsection{The impact of malicious user number (RQ4)}\label{sec_xi}

\textbf{PSMU with Extremely Fewer Malicious Users.}
In Fig.~\ref{fig_att_base}, we already show that PSMU can be effective with $\xi=0.1\%$, i.e. $6$ and $16$ malicious users on ML and AZ datasets respectively.
In this part, we further investigate if PSMU can promote target items with extremely fewer malicious users, since fewer malicious users represent that the cost of launching such an attack is less, which further reveals the severe threats.
Fig.~\ref{fig_att_proportion} shows PSMU's performance with a different number of malicious users.
In Fig.~\ref{fig_att_proportion_ml}, only employing $3$ and $2$ malicious users, PSMU achieves $1.0$ ER@5 on ML with Fed-NCF and Fed-LightGCN, respectively.
In Fig.~\ref{fig_att_proportion_az}, $12$ and $4$ malicious users can help PSMU promote target items to all clients.

\textbf{HiCS with More Malicious Users.} 
When more malicious users are employed in an attack, the defense will be more challenging. 
In Table~\ref{tb_defense_proportion}, we compare HiCS with $\ell_{2}$ clipping under increasing $\xi$. 
For each value of $\xi$, we report the highest ER@5 value PSMU achieved during the whole training process under defense protection.
Since $\ell_{2}$ clipping is one of the subcomponents of HiCS, this comparison can also indicate the effectiveness of adaptive clipping with sparsified updating.
As observed in Table~\ref{tb_defense_proportion}, $\ell_{2}$ clipping is only effective in Fed-NCF when $\xi=0.1\%$.
After being equipped with sparsified updating, the defense can be effective in all cases when $\xi=0.1\%$, which shows the effectiveness of sparsified updating.
However, with more malicious users, both above methods cannot defend against PSMU any longer.
Only HiCS keeps consistent effectiveness even when malicious user proportion increased to even $10\%$ proportion.

\section{Conclusion}
In this paper, we propose an effective poisoning attack, PSMU, which attacks federated recommender systems (FedRecs) without prior knowledge and with fewer malicious users and fewer epochs, revealing the vulnerability of FedRecs to gradient poisoning attacks.
Then, we take the first step to explore how to defend against gradient poisoning attacks in FedRecs and propose a novel defense method named HiCS.
To show the effectiveness of our attack and defense methods, we conduct extensive experiments with two widely used FedRecs on two real-world recommendation datasets.
The comparison with state-of-the-art attacks and widely used defense baselines demonstrates the superiority of both PSMU and HiCS.

\begin{acks}
  This work is supported by Australian Research Council Future Fellowship (Grant No. FT210100624), Discovery Project (Grant No. DP190101985), and Discovery Early Career Research Award (Grant No. DE200101465).
\end{acks}

\bibliographystyle{ACM-Reference-Format}
\bibliography{sample-base}










\end{document}